\documentclass[
aps,
pra,
superscriptaddress,
12pt,
reprint,
nofootinbib
]{revtex4-1}

\usepackage[utf8]{inputenc}
\usepackage[T1]{fontenc}
\usepackage[english]{babel}

\usepackage{graphicx}

\usepackage{amsmath}
\usepackage{amssymb}
\usepackage{mathtools}

\usepackage[colorlinks=true, allcolors=blue]{hyperref}
\usepackage[all]{hypcap}  

\usepackage{braket}
\usepackage{xcolor}
\usepackage[normalem]{ulem}
\usepackage{nicefrac}

\renewcommand{\v}[1]{\boldsymbol{#1}}
\newcommand{\Rb}{$^{87}\textrm{Rb}$ }

\begin{document}
\title{Electric field-induced wave-packet dynamics and geometrical rearrangement of trilobite Rydberg molecules}
\date{\today}
\author{Frederic Hummel}
\email{frederic.hummel@physnet.uni-hamburg.de}
\affiliation{Zentrum für Optische Quantentechnologien, Fachbereich Physik, Universität Hamburg, Luruper Chaussee 149, 22761 Hamburg, Germany}
\author{Kevin Keiler}
\affiliation{Zentrum für Optische Quantentechnologien, Fachbereich Physik, Universität Hamburg, Luruper Chaussee 149, 22761 Hamburg, Germany}
\author{Peter Schmelcher}
\affiliation{Zentrum für Optische Quantentechnologien, Fachbereich Physik, Universität Hamburg, Luruper Chaussee 149, 22761 Hamburg, Germany}
\affiliation{The Hamburg Centre for Ultrafast Imaging, Universität Hamburg, Luruper Chaussee 149, 22761 Hamburg, Germany}

\begin{abstract}
	We investigate the quantum dynamics of ultra-long-range trilobite molecules exposed to homogeneous electric fields. A trilobite molecule consists of a Rydberg atom and a ground-state atom, which is trapped at large internuclear distances in an oscillatory potential due to scattering of the Rydberg electron off the ground-state atom. Within the Born-Oppenheimer approximation, we derive an analytic expression for the two-dimensional adiabatic electronic potential energy surface in weak electric fields valid up to 500 V/m. This is used to unravel the molecular quantum dynamics employing the Multi-Configurational Time-Dependent Hartree method. Quenches of the electric field are performed to trigger the wave packet dynamics including the case of field inversion. Depending on the initial wave packet, we observe radial intra-well and inter-well oscillations as well as angular oscillations and rotations of the respective one-body probability densities. Opportunities to control the molecular configuration are identified, a specific example being the possibility to superimpose different molecular bond lengths by a series of periodic quenches of the electric field.
\end{abstract}

\maketitle

\section{Introduction}

	Trilobite Rydberg molecules are a prototype of an exotic molecular species called ultra-long-range Rydberg molecules (ULRM). The binding mechanism of ULRM is based on the low-energy scattering of the highly excited valence electron of a Rydberg atom off a ground-state atom. These molecules differ from conventional molecules with covalent or ionic bonds by their gigantic bond lengths of several hundreds or even thousands of angstroms and correspondingly huge dipole moments of up to a few kilo debye, e.g.~in the case of trilobite molecules. Qualitatively, ULRM can be divided into two classes according to the orbital angular momentum $l$ of the Rydberg electron: Non-polar molecules for $l\leq3$, when the electronic state is energetically split off the quasi-degenerate hydrogen-like manifold due to a non-integer quantum defect and polar molecules with $l>3$ such as the trilobite, in which the Rydberg electron is in a superposition of high angular momentum states.
	
	ULRM have first been explored theoretically in the year 2000 \cite{Greene2000} and discovered experimentally in 2008 \cite{Bendkowsky2009}. They have been realized experimentally in cold or ultra-cold ensembles of the atomic species rubidium \cite{Bendkowsky2010,Butscher2010,Butscher2011,Li2011,Bellos2013,Kleinbach2017}, cesium \cite{Tallant2012,Booth2015}, strontium \cite{DeSalvo2015,Camargo2016}, and even in hetero-nuclear gases \cite{Whalen2020,Peper2021}. Their binding energies are on the order of a few GHz in the case of polar molecules and of a few MHz for non-polar molecules. Several extensions of the theoretical modeling of ULRM alongside with the increasing resolution of the experiments have been developed to accommodate for scattering of higher partial waves \cite{Hamilton2002,Chibisov2002}, electronic fine and hyperfine structure \cite{Anderson2014th}, and spin-orbit coupling \cite{Khuskivadze2002,Markson2016a,Eiles2017,Deiss2020}. This led to applications of ULRM for the characterization of low-energy electron-atom collisions \cite{Anderson2014exp,Sassmannshausen2015,Boettcher2016,MacLennan2019,Engel2019} and for probing spatial correlations in the underlying atomic gases \cite{Manthey2015,Whalen2019,Whalen2019a}. Their molecular properties such as dipole moments or alignment can be controlled via weak external electric and magnetic fields \cite{Lesanovsky2006,Kurz2013,Krupp2014,Gaj2015,Niederprum2016,Hummel2018,Hummel2019}. Recently, ULRM have been proposed as candidates for the production of ultra-cold negative ions \cite{Hummel2020, Peper2020} and other classes of exotic molecules via ion-pair dressing \cite{Giannakeas2020b} and Rydberg dressing \cite{Wang2020}. 
	
	So far, all studies of ULRM both theoretically and experimentally have been focusing on the electronic and vibrational structure. In this work, we explore the quantum dynamics and wave packet evolution of trilobite molecules exposed to a weak electric field. The presence of the field breaks the spherical symmetry of the molecule which leads effectively to a two-dimensional adiabatic potential energy surface. For the nuclear quantum dynamics, we employ the Multi-Configurational Time-Dependent Hartree (MCTDH) method \cite{Meyer1990, Beck2000, Meyer2003, Meyer2012}, an ab initio wave packet propagation method that combines the benefits of numerical precision and low computational cost by variationally optimizing both basis functions and coefficients at each time step. 
	
	\begin{figure*}
		\centering
		\includegraphics[width=0.95\textwidth]{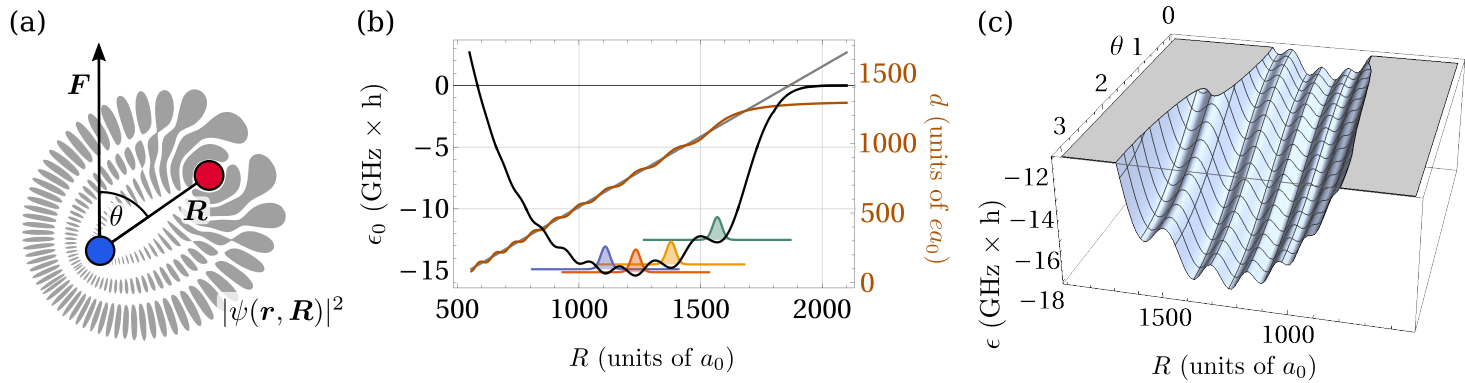}
		\caption{(a) Illustration of a trilobite molecule in an electric field. The Rydberg electronic density (gray) ties the ionic core (blue) and the ground-state atom (red) together along $\v{R}$. $\theta$ is the angle relative to the electric field axis $\v{F}$. (b) The field-free adiabatic potential $\epsilon_0(R)$ (black) along with the dipole moment $d(R)$ (brown), a classical dipole $r-n^2/2$ for comparison (gray), and selected vibrational levels (colored). (c) The two-dimensional adiabatic potential $\epsilon(R,\theta)$ for the field strength $F=200\,\mathrm{V/m}$. \label{fig:scheme}}
	\end{figure*}
	
	To trigger the dynamics, we focus on different quenches of the electric field. In our first scenario, the ULRM is initialized in an eigenstate of the field-free configuration with an isotropic angular distribution. Following a sudden switch on of the field, we observe a correlated wave packet dynamics classified as intra-well oscillations on different time-scales. Secondly, the ULRM is initialized in an eigenstate for a specific field strength $F_0$ such that the angular density is localized and the field is quenched to a different field strength $F$ with opposite direction. For strong fields, additional inter-well oscillations occur and the molecule exists in a superposition of different geometries. Finally, we explore the case of a periodic sequence of quenches. This allows us to prepare the molecule in equally mixed radial eigenstates with a distinct radial localization in different potential wells deviating by hundreds of angstroms. 

	Our work is structured as follows. In section \ref{sec:meth}, we provide the underlying methodology covering the electronic structure, the field impact, and the nuclear dynamics. Our numerical results are presented in section \ref{sec:res} addressing the cases of single quenches in subsections \ref{sec:iso} and \ref{sec:loc} and periodic sequences of quenches in subsection \ref{sec:perio}. Section \ref{sec:conc} contains the conclusions and a brief outlook.
	
\section{Methodology} \label{sec:meth}

\subsection{Electronic structure \label{sec:elec}}

	Let us consider two \Rb atoms, one of which is in a Rydberg state and the second one is in its ground state. The ionic core of the Rydberg atom is located at the coordinate origin and the excited electron at position $\v{r}$, while the position of the ground-state atom is $\v{R}$. According to the Born-Oppenheimer approximation, the adiabatic electronic Hamiltonian is given by $H_e=H_c+V$, where $H_c$ describes the Rydberg electron in the potential of the ionic core and $V$ models the interaction between the Rydberg electron and the ground-state atom. The low angular momentum states $l<3$ are energetically detuned from a degenerate manifold of high angular momentum states. The latter are well described by hydrogenic wave functions $\phi_{nlm}(\v{r})$ with energies $-1/(2n^2)$ (in atomic units), where $l\geq 3$ and $n$ and $m$ are the principle and magnetic quantum numbers, respectively. 
	
	The interaction of the Rydberg electron with the ground-state atom is modeled by a scattering pseudo potential \cite{Fermi1934}
	\begin{equation}
		V=2\pi a(k)\delta(\v{r}-\v{R}),
	\end{equation}
	where $a(k)$ is the energy-dependent scattering length obtained via modified effective range theory \cite{Spruch1960, OMalley1961} $a(k)=a(0)+\pi \alpha k/3$, with the electron's wave number $k$, polarizability $\alpha=319.2$ \cite{Arimondo1977}, and zero-energy scattering length $a(0)=-16.1$ \cite{Bahrim2001}. The wave number is determined in a semiclassical approximation via $k^2/2-1/R=-1/(2n^2)$. 
	
	This simplistic model Hamiltonian captures the essential features of trilobite Rydberg molecules \cite{Greene2000,Bendkowsky2009} and is sufficient for the purpose of the present work: A quantum dynamical study of ULRM. Quantitative corrections originate from $p$-wave scattering \cite{Hamilton2002, Chibisov2002}, fine structure of the Rydberg atom and hyperfine structure of the ground-state atom \cite{Anderson2014th}, as well as spin-spin and spin-orbit interactions \cite{Markson2016a, Eiles2017}. The Hamiltonian $H_e$ can be solved analytically by degenerate perturbation theory \cite{Greene2000}. Alternative approaches to the electronic structure rely on Green's function methods \cite{Khuskivadze2002, Fey2015, Tarana2016} or local frame transformation \cite{Giannakeas2020a}. The result is the (unnormalized) trilobite state
	\begin{equation}
		|\psi(\v{r},\v{R})|^2 = \sum_{lm} \phi_{nlm}^\star(\v{R}) \phi_{nlm}(\v{r}),
	\end{equation}
	and the corresponding trilobite potential
	\begin{equation}
		\epsilon_0(R) = 2\pi a(k(R)) |\psi(\v{R},\v{R})|^2, \label{eq:free}
	\end{equation}
	which depends only on the internuclear distance $R$ and is discussed in section \ref{sec:elec} and shown in figure \ref{fig:scheme} (b).
	
\subsection{Impact of an electric field \label{sec:field}}	

	We consider an electric field parallel to the $z$ axis of our coordinate system. This leads to an additional term in the Hamiltonian, which can be evaluated in perturbation theory to
	\begin{eqnarray}
		\braket{\psi|\v{F}\v{r}|\psi} = F d(R) \cos\theta,
	\end{eqnarray}
	where $F$ is the electric field strength, $\theta$ is the relative angle between the electric field and the internuclear axis, and $d(R)$ is the dipole moment function of the trilobite state as seen in figures \ref{fig:scheme} (a) and (b), respectively. This approximation is valid as long as the interaction of the Rydberg electron with the ground-state atom dominates the electric field interaction. The latter holds, if the Stark splitting of the degenerate hydrogenic manifold of high angular momentum states, which scales as $Fn^2$, remains small compared to the potential depth of the trilobite's adiabatic electronic potential energy surface, which scales as $n^{-3}$. This means that the critical field strength decreases rapidly as $n^{-5}$. However, for $n=30$, we have ensured that the perturbative results are in excellent agreement with the results of an exact diagonalization of the complete electronic Hamiltonian in a finite basis set up to field strength of $F=500\,\mathrm{V/m}$. The deviations due to the non-perturbative contribution to the potential are less than 1\% for the internuclear distances covered in the dynamics.
	
	In total, we obtain a two-dimensional potential energy surface, 
	\begin{equation}
		\epsilon(R,\theta) = \epsilon_0(R) + F d(R) \cos\theta, \label{eq:field}
	\end{equation}
	which depends on the internuclear separation $R$ of the ionic core of the Rydberg atom and the ground-state atom and the relative angle $\theta$ between the internuclear axis and the electric field. The sign of $F$ determines the field direction. This potential is discussed in section \ref{sec:elec} and shown in figure \ref{fig:scheme} (c).
	
\subsection{Nuclear dynamics \label{sec:nuc}}

	To analyze the nuclear dynamics, we have to solve the corresponding time-dependent Schrödinger equation
	\begin{eqnarray}
		i \frac{\partial}{\partial t} \chi(R,\theta,t) &=& H_n \chi(R,\theta,t), \label{eq:tds} \\
		H_n &=& -\frac{1}{2\mu}\left(\Delta_R+\Delta_\theta\right) + \epsilon(R,\theta), \\
		\Delta_R &=& \frac{\partial^2}{\partial R^2} + \frac{2}{R} \frac{\partial}{\partial R} \label{eq:dr} \\
		\Delta_\theta &=& \frac{1}{R^2} \left( \frac{\partial^2}{\partial \theta^2} + \cot \theta \frac{\partial}{\partial \theta} \right) \label{eq:dth}
	\end{eqnarray}
	where $\mu=m/2$ and $m$ is the atomic mass of $^{87}\textrm{Rb}$. Due to the cylindrical symmetry of the system, the polar angle $\phi$ is a cyclic coordinate and the corresponding  magnetic quantum number $M$ is a conserved quantity. Here, we restrict ourselves to the case $M=0$. To solve eq.~(\ref{eq:tds}), we employ the Multi-Configurational Time-dependent Hartree (MCTDH) approach \cite{Meyer1990, Beck2000, Meyer2003, Meyer2012}. In the following, we briefly describe the underlying method. More detailed descriptions are provided for example in references \cite{Beck2000, Meyer2012}. 
	
	MCTDH is an ab initio tool for propagating wave packets in high dimensional spaces. Within this approach the time-dependent nuclear wave function $\chi(R,\theta,t)$ of our trilobite molecule in an electric field is expressed as
	\begin{equation}
		\chi(R,\theta,t) = \sum_{i_1=1}^{n_1} \sum_{i_2=1}^{n_2} \tilde c_{i_1,i_2}(t) \tilde\varphi_{i_1}^{(1)}(R,t) \tilde\varphi_{i_2}^{(2)}(\theta,t), \label{eq:mctdh}
	\end{equation}
	where $\tilde c_{i_1,i_2}(t)$ is a time-dependent coefficient, $\tilde\varphi_{i_d}^{(d)}(q_d,t)$ is the $i_d$-th time-dependent single-particle function (SPF) of the $d$-th degree of freedom $q_d\in\{R,\theta\}$, and $n_d$ is the number of single particle functions employed for the $d$-th degree of freedom. For the case of two degrees of freedom, by employing the Schmidt decomposition of the two orthonormal basis sets represented by the SPFs, we can choose $n_1=n_2 \equiv N$. The square matrix of coefficients $\tilde C(t)$ can now be diagonalized and the SPFs are transformed accordingly such that we obtain
	\begin{equation}
		\chi(R,\theta,t) = \sum_i^N c_i(t) \varphi_i^{(1)}(R,t) \varphi_i^{(2)}(\theta,t). \label{eq:mctdh_final}
	\end{equation}
	The key idea of the MCTDH method is to keep the number of necessary SPFs small by employing variationally optimized coefficients $c_{i}(t)$ and basis functions $\varphi_i^{(d)}(q_d,t)$ obtained from the Dirac-Frenckel variational principle $\braket{\delta\chi|i\partial_t-H_n|\chi}=0$ \cite{Dirac1930,Frenkel1934} and requiring orthonormality of the SPFs $\braket{\varphi_i|\varphi_j}=\delta_{ij}$ and $\braket{\varphi_i|\dot\varphi_j}=0$. This ansatz allows a sizable reduction of the computational cost compared to e.g.~a corresponding approach where only the coefficients are time-dependent. 
	
	The SPFs are normalized as 
	\begin{eqnarray}
	\int\mathrm{d}R\,\left|\varphi_i^{(1)}(R,t)\right|^2&=&1, \\ \int\mathrm{d}\theta\sin\theta\,\left|\varphi_i^{(2)}(\theta,t)\right|^2&=&1
	\end{eqnarray}
	and represented on a grid by a discrete variable representation (DVR) \cite{Harris1965}. In our case, the natural choices are the sine DVR for the radial degree of freedom, where we use 712 grid points, and the Legendre DVR for the angular degree of freedom, where we use 601 grid points. 
	
	In this study, we employ two different types of initial states for the wave packet propagation. Firstly, we consider field-free molecular configurations. In this case, the degrees of freedom $R$ and $\theta$ decouple, since the electronic potential does not depend on the angle $\theta$. MCTDH provides a build-in function to obtain the field-free vibrational eigenstates $\xi_{i_d}^{(d)}(q_d)$ of the one-body Hamiltonians
	\begin{eqnarray}
	H_R &=& -\frac{1}{2\mu} \Delta_R + \epsilon_0(R), \\
	H_\theta &=& -\frac{1}{2\mu} \Delta_\theta, \label{eq:theta}
	\end{eqnarray}
	using the Lanczos algorithm, for $\Delta_R$ and $\Delta_\theta$ as defined in equations (\ref{eq:dr}) and (\ref{eq:dth}), respectively. The solutions of eq.~(\ref{eq:theta}) are the Legendre polynomials. Secondly, we consider the vibrational ground state in non-zero electric fields. To this end, we employ the relaxation method of the MCTDH method that uses imaginary-time evolution of the Hamiltonian $H_n$. 
	
	The number of SPFs necessary in order to retrieve converged results of the propagation depends strongly on the field strength under consideration, the chosen initial state, and the quench protocol. We consider a calculation to be converged in this context, if the occupation of the $N$-th SPF remains sufficiently close to zero, i.e.~below 0.1\%, and the occupations $c_i$ decrease exponentially. For propagation in fields up to $F=50\,\mathrm{V/m}$, we obtain converged results for $N=8$, for fields up to $F=400\,\mathrm{V/m}$, we employ $N=32$. 
	
\section{Results and discussion} \label{sec:res}

\subsection{Adiabatic potential energy surfaces \label{sec:pot}}

	Before we investigate the nuclear dynamics of the trilobite molecule in an electric field, let us briefly review the main properties of the underlying adiabatic electronic potential energy surface. In the field-free case, the multi-well structure of the trilobite potential $\epsilon_0(R)$ (eq.~(\ref{eq:free})) allows for localized vibrational states at the distinct radial positions of the potential wells. For the example case of $n=30$, five such wells are of relevance as can be seen in figure \ref{fig:scheme} (b). The four outermost wells host a Gaussian-shaped state (shown in colors in the same figure), while the state dominantly localizing in the inner well around $R=1000\,a_0$ (not shown in figure \ref{fig:scheme} (b)) possesses a non-zero probability amplitude in the outer wells. Additionally, the three potential wells between $1100\,a_0<R<1400\,a_0$ each host an excited localized state with a single node. Other vibrational states are typically delocalized over two or more wells. Counting from the outermost well at large internuclear distances, the third well hosts the global potential minimum and the vibrational ground-state $\xi_{0}^{(1)}$ (red), where the upper index refers to the radial degree of freedom. The first excited state $\xi_{1}^{(1)}$ (blue) localizes in the neighboring well at smaller internuclear distance, while the second excited state $\xi_{2}^{(1)}$ exhibits a single node and localizes in the same well as the ground-state. 
	
	The dipole moment function of the trilobite molecule $d(R)$ increases with increasing $R$. This can be understood by the fact that the potential energy of the molecule is minimized, when the Rydberg electron localizes density at the position of the perturbing ground-state atom. Thus, the distance of the ground-state atom to the ionic core of the Rydberg atom determines the separation of the positive charge of the ionic core and the negative charge of the electron. For large internuclear distances $R>2n^2=1800\,a_0$, we approach the dissociation limit and the dipole moment saturates which is shown in figure \ref{fig:scheme} (b) in brown. 
	
	Due to the large dipole moment of the trilobite molecule, the relative depth of the potential wells can be tuned by rather weak electric field strengths \cite{Kurz2013}. In an arbitrary non-zero field, the equilibrium configuration is $\theta=\pi$ (compare figure \ref{fig:scheme} (c)). At a field strength of $F\approx325\,\mathrm{V/m}$, the second outermost potential well becomes as low in energy as the third one. Therefore, for larger fields, the second outermost well hosts the global potential minimum in the configuration $\theta=\pi$, such that the vibrational ground state is shifted to the adjacent well. Oppositely, for the configuration $\theta=0$, the inner wells increase in depth relative to the outer wells. At a field strength stronger than $F\approx110\,\mathrm{V/m}$, the fourth outermost well becomes lower in energy as the third. This highlights the relevance of the angular dynamics for the overall radial configuration of the molecule: Rotation of a molecule in an electric field may induce oscillations between different radial configurations.
	
	\subsection{Quench dynamics for an isotropic initial state \label{sec:iso}}

	\begin{figure*}
		\centering
		\includegraphics[width=0.9\textwidth]{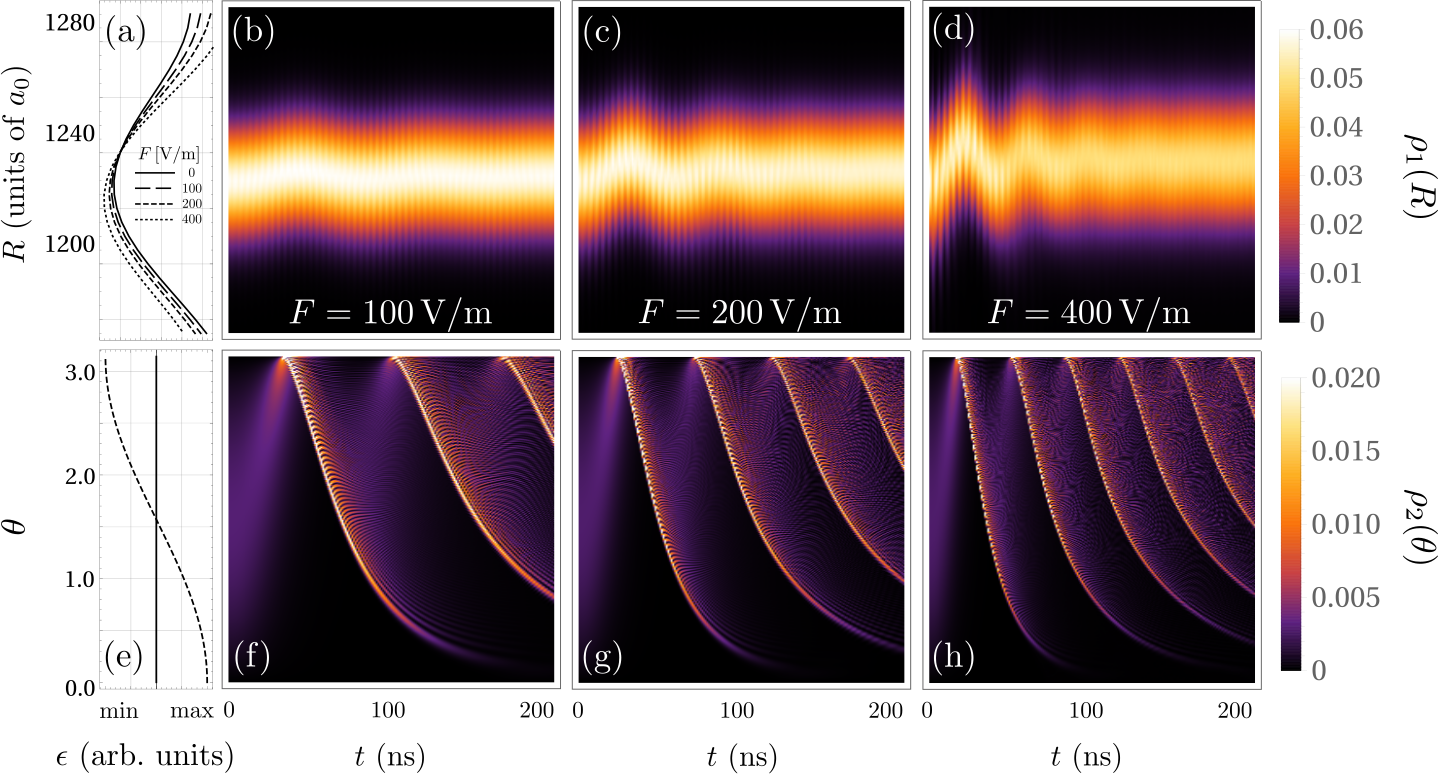}
		\caption{Molecular wave packet dynamics after a sudden quench of the electric field, initialized from the isotropic, field-free ground state. (a) Dynamically relevant section of the radial potential before (solid) and after the quench to different field strengths $F$. (b)-(d) Temporal evolution of the radial density $\rho_1(R)$ corresponding to the fields. (e) The angular potential before (solid) and after (dashed) the quench. (f)-(h) Temporal evolution of the corresponding angular density $\rho_2(\theta)$. Two time scales of radial oscillations are apparent, the slower of which is correlated to the angular oscillations and depends on the field strength $F$. \label{fig:iso}}
	\end{figure*}
	
	In a first series of simulations, we prepare the molecule in the isotropic, field-free ground state and switch on the electric field at $t=0$ i.e.~the potential is quenched from $\epsilon_0(R)$ (eq.~(\ref{eq:free})) to $\epsilon(R,\theta)$ (eq.~(\ref{eq:field})). Figure \ref{fig:iso} shows both the radial (b)-(d) and the angular (f)-(g) dynamics of the molecule, while (a) and (e) show the dynamically relevant sections of the radial and angular potential, respectively, as a guide for the eye. Column-wise, the figure shows the dynamics for different field strength ranging from $F=100-400\,\mathrm{V/m}$, each for an evolution time of 200 ns. The radial one-body probability density (OBD) 
	\begin{equation}
		\rho_1(R,t) = \sum_i \left|c_i(t) \varphi_i^{(1)}(R,t)\right|^2
	\end{equation}	
	oscillates inside the initially populated potential well. Hence, we classify the dynamics as \emph{intra-well} oscillations. Furthermore, the radial OBD reveals two distinct time scales for radial dynamics: A high-frequency oscillation with a period of few ns and a slow oscillation with a period of $30-100\,\mathrm{ns}$ depending on the field strength. 

	\begin{figure}
		\centering
		\includegraphics[width=0.32\textwidth]{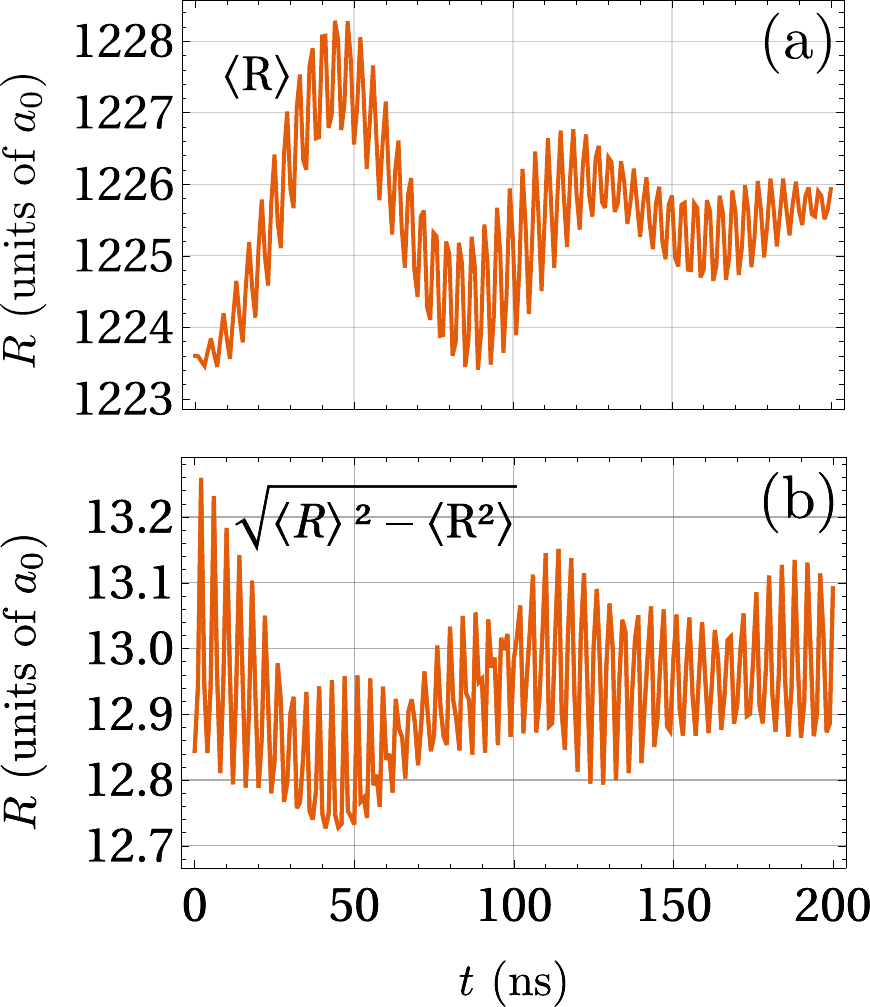}
		\caption{Mean value (a) and variance (b) of the radial one-body density after a quench of the electric field to $F=100\,\mathrm{V/m}$ initialized from the isotropic, field-free ground state. This corresponds to the density shown in figure \ref{fig:iso} (b) and highlights the presence of two dipole-oscillation modes accompanied by breathing oscillations. \label{fig:mean}}
	\end{figure}	

	This is highlighted in figure \ref{fig:mean}, where the mean value (a) and the variance (b) of the radial OBD is shown in the case of $F=100\,\mathrm{V/m}$ (compare figure \ref{fig:iso} (b)). Shortly after the quench, the high-frequency oscillation is dominated by breathing (oscillation of the variance), however, after a few nano seconds, this is accompanied by dipole-type oscillations (oscillation of the mean). The period of a few ns is independent of the field strength, however, the corresponding amplitude slightly increases for stronger fields (not shown in figure \ref{fig:mean}, compare figure \ref{fig:iso} (b)-(d)). This can be attributed to a small field-dependent displacement of the potential well, which is shown in \ref{fig:iso} (a). In this figure, the absolute energetic position of the wells at different field strengths is neglected and only the relative well depth and radial position is shown.
	
	The slow oscillation with a period of $30-100\,\mathrm{ns}$ is dominantly of dipole character. Its period corresponds to the period of the angular dynamics (compare figure \ref{fig:iso} (f)-(h)): From the isotropic initial state, the angular OBD 
	\begin{equation}
		\rho_2(\theta,t) = \sum_i \sin\theta\left|c_i(t) \varphi_i^{(2)}(\theta,t)\right|^2
	\end{equation}	
	maximizes close to the equilibrium position obtaining a very steep slope around $\theta=\pi$. Afterwards, an oscillatory behavior along $\theta$ emerges with a period that decreases for increasing field strength. This dynamics can be compared to a classical pendulum, where the initial displacement does not influence the period of the oscillation. Therefore, the isotropic initial state $\xi_{0}^{(2)}$ can be understood as a superposition of all possible displacements which converge at the equilibrium position $\theta=\pi$ after the same evolution time. This picture, however, only holds for the first oscillation ($t<100\,\mathrm{ns}$ in figure \ref{fig:iso} (f)). Afterwards, we find that part of the wave packet performs a large amplitude oscillation, while significant parts of it already maximize again around the equilibrium position (compare for example figure \ref{fig:iso} (f) around $t=100\,\mathrm{ns}$). The oscillation period scales as $1/\sqrt{F}$ corresponding to the classical scaling of a dipole in an external electric field. The radial OBD oscillates with the same period and the radial displacement is maximal when the angular wave packet localizes at the equilibrium position. Here, the wave packet's velocity and the corresponding centrifugal force are largest. Consequently, the amplitude of the oscillation increases for larger fields and is therefore most prominent in figure \ref{fig:iso} (d). For long evolution times the oscillations are damped corresponding to the broadening distribution of the angular OBD.
	
\subsection{Quench dynamics for a localized initial state \label{sec:loc}}

	\begin{figure*}
		\centering
		\includegraphics[width=0.9\textwidth]{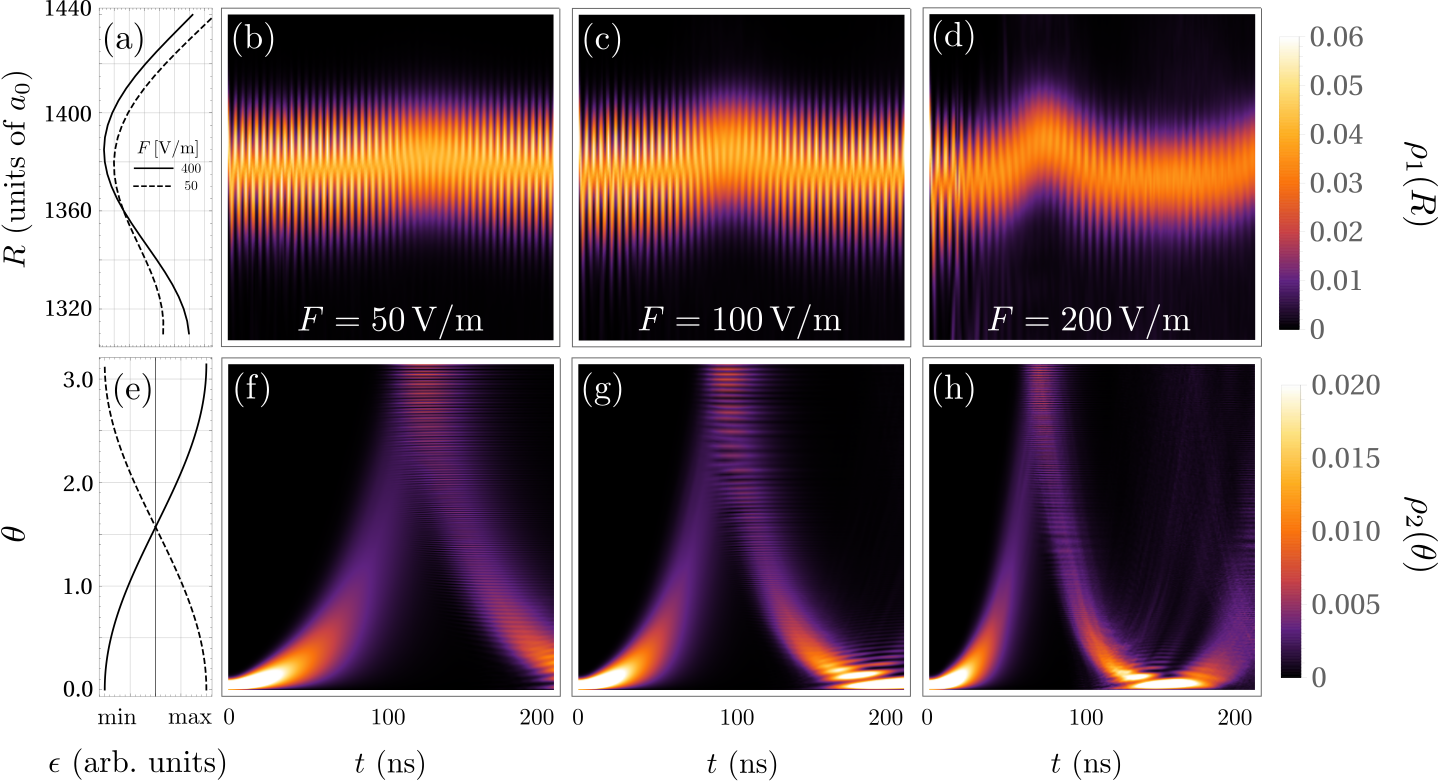}
		\caption{Molecular wave packet dynamics after a sudden inversion of the electric field direction initialized from the localized, vibrational ground state at $F_0=-400\,\mathrm{V/m}$. (a) Dynamically relevant section of the radial potential before (solid) and after the quench (dashed) in the case of $F=50\,\mathrm{V/m}$. (b)-(d) Temporal evolution of the radial radial density $\rho_1(R)$ for different field strengths $F$. (e) The angular potential before (solid) and after (dashed) the quench. (f)-(h) Temporal evolution of the corresponding angular density $\rho_2(\theta)$. The angular dynamics correspond to a rotation. \label{fig:loc}}
	\end{figure*}

	Opposite to the quench discussed in the previous section, here, we investigate the molecular dynamics initialized from the vibrational ground state in the presence of an electric field $F_0=-400\,\mathrm{V/m}$. The quench consists of an inversion of the electric field direction. Our particular choice has two effects: Firstly, switching the sign of the electric field direction shifts the equilibrium position from $\theta=0$ (pre-quench) to $\theta=\pi$ (post-quench) and, secondly, choosing a large pre-quench field strength shifts the vibrational ground state to the adjacent outer potential well as described in section \ref{sec:pot}.
	
	Figure \ref{fig:loc} is structured similarly to figure \ref{fig:iso} and shows the dynamically relevant sections of the radial (a) and angular (e) potentials as well as the temporal evolution of the radial (b)-(d) and angular OBD (f)-(g) for field strengths from $F=50$ to $200\,\mathrm{V/m}$. We find that for this quench protocol, the radial intra-well oscillations are more prominent and of dipole character. Again, there are two distinct time-scales: A high-frequency oscillation with a period of few ns and a slow oscillation with a field-dependent period. The period of the high-frequency oscillation does not depend on the field strength, however, its amplitude slightly increases with increasing field strength. In a weak field of $F=50\,\mathrm{V/m}$, the angular OBD (f) disperses from its initial position around $\theta=0$ and passes through the new equilibrium position at $\theta=\pi$ after approximately $t=120\,\mathrm{ns}$. At that time, an interference pattern is visible, which we attribute to the molecule's fast radial oscillations, which in a classical picture leads to different rotational constants and consequently slightly differing arrival times at $\theta=\pi$. Afterwards, the wave packet travels back to its initial position, where it arrives at approximately $t=250\,\mathrm{ns}$ (not shown in the figure). This behavior can be interpreted as a molecular rotation in contrast to the oscillation discussed in the previous section, where not the entire range of $\theta$ is covered. The rotation period decreases with increasing field. When the angular OBD has performed one half of its rotation, we find a displacement of the radial OBD to larger internuclear distances. This corresponds to the dynamics described in the previous section and can be attributed to the centrifugal force induced by the rotation, which becomes largest for the highest speed of the wave packet occuring at $\theta=\pi$.
		
	\begin{figure*}
		\centering
		\includegraphics[width=0.75\textwidth]{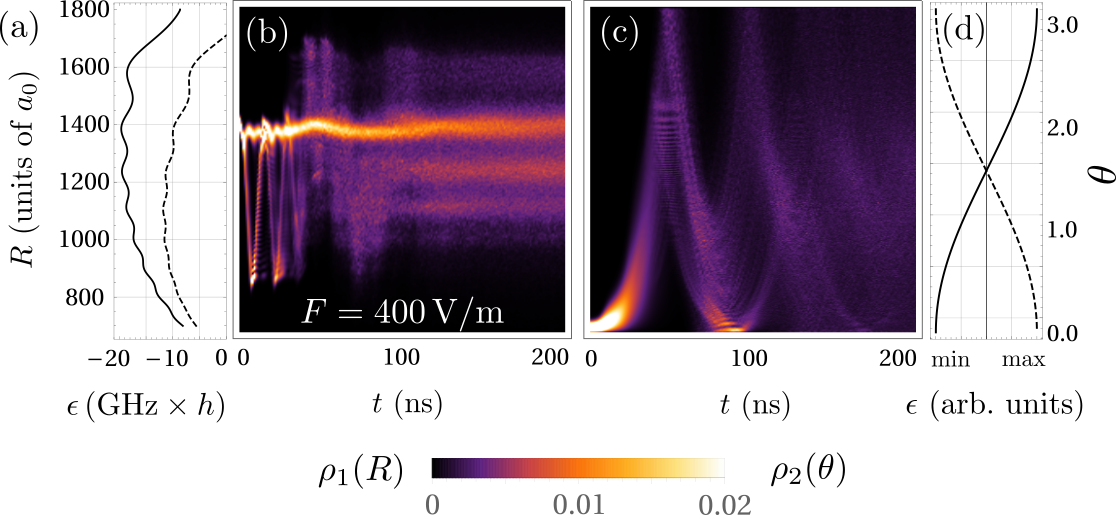}
		\caption{Molecular wave packet dynamics after a sudden inversion of the electric field direction initialized from the localized, vibrational ground state at $F_0=-400\,\mathrm{V/m}$. (a) The dynamically relevant cuts of the radial potential at $\theta=0$ before (solid) and after (dashed) the quench (or vice versa at $\theta=\pi$). (b) Temporal evolution of the radial and (c) angular densities. (d) The angular potential before (solid) and after (dashed) the quench. For this field strength, radial, large-amplitude, inter-well oscillations occur and correspondingly, the angular rotation blurs out after approximately half its period. \label{fig:loc400}}
	\end{figure*}
	
	The overall structure of the molecular dynamics changes drastically in a stronger field. Figure \ref{fig:loc400} shows the radial (b) and angular (c) OBD in a field of $F=400\,\mathrm{V/m}$ along with the dynamically relevant sections of the radial (a) and angular (d) potentials before (solid) and after (dashed) the quench. Due to the symmetry $F=-F_0$, in (a), the radial potentials correspond additionally to different configurations $\theta=\pi$ (dashed) and $\theta=0$ (solid) after the quench (or inversely before the quench). We observe oscillations across many potential wells which we classify as \emph{inter-well} oscillations. At short evolution times, the angular OBD exhibits the same behavior as shown in figure \ref{fig:loc} (e)-(g), where the angular wave packet rotates along $\theta$. However, after half a rotation at approximately $t=50\,\mathrm{ns}$, the wave packet is blurred out and after approximately $t=150\,\mathrm{ns}$, an almost equal superposition of angles is reached. This can be attributed to the vastly varying arclengths associated to to radially oscillating wave packet such that a coherent rotation is not possible. Correspondingly, the radial inter-well dynamics also effectively blurs out after the angular wave packet has performed half a rotation. The initial radial wave packet is localized at approximately $R=1400\,a_0$, which corresponds to the outer potential well. In the first $50\,\mathrm{ns}$, it performs three radial, large-amplitude oscillations covering a range of $600\,a_0$, before the oscillation blurs out. This corresponds to the strong tilt that the potential experiences after changing the direction of the electric field. After an evolution time of $t=100\,\mathrm{ns}$, the radial OBD is widely spread between $1000\,a_0<R<1600\,a_0$.

\subsection{Periodic quenches \label{sec:perio}}

	\begin{figure*}
		\centering
		\includegraphics[width=1.0\textwidth]{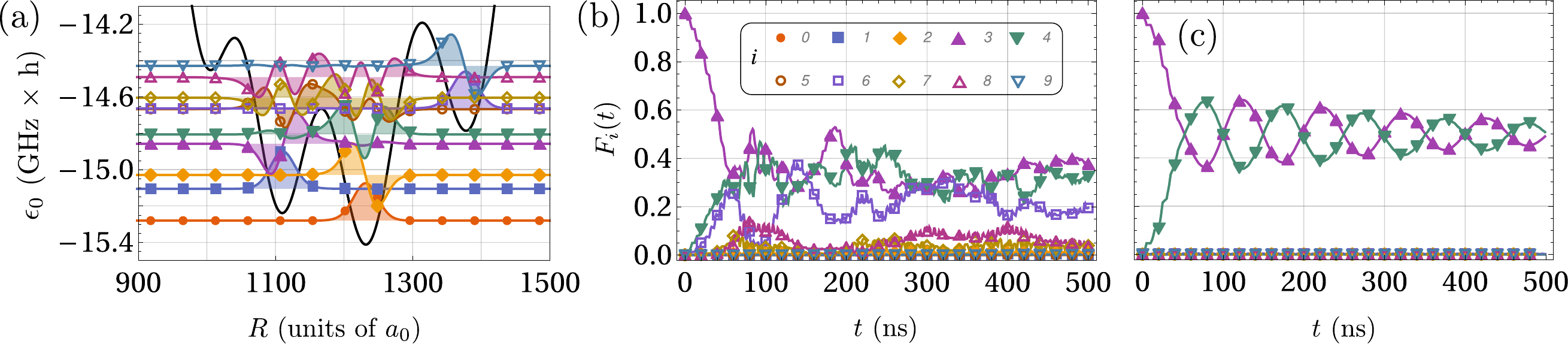}
		\caption{Analysis of the radial components of the initial wave function. (a) The field-free trilobite potential (solid) with its vibrational eigenstates (symbol and color). (b)-(c) Fidelities $F_i$ of the radial components with these eigenstates (same symbol and color) for a waiting time between quenches of (b) $T=3\,\mathrm{ns}$ and (c) $T=5\,\mathrm{ns}$. The waiting time strongly influences the number of contributing states. In (c) the quenches are discontinued after $t=50\,\mathrm{ns}$ and coherent oscillations follow. \label{fig:perio}}
	\end{figure*}
		
	In the second series of simulations, we switch the field on and off periodically. In other words, we apply a sequence of quenches back and forth between the potentials $\epsilon_0(R)$ and $\epsilon(R,\theta)$. Here, the waiting time $T$ between quenches plays a crucial role. Exemplarily, we consider a radially excited initial state from the isotropic, field-free configuration to initialize the dynamics. Specifically, we consider the radial eigenstate $\xi_3^{(1)}$ that localizes in an inner well of the trilobite and exhibits a single radial node. The field is switched periodically between $F=0$ and $50\,\mathrm{V/m}$. 	
	
	Figure \ref{fig:perio} shows a state analysis for the waiting times $T=3$ and $5\,\mathrm{ns}$. To this end, we focus on the radial components of the nuclear wave function, i.e.~the radial OBD $\rho_1(R,t)$. Figure \ref{fig:perio} (a) shows a magnification of the relevant wells of the field-free trilobite potential $\epsilon_0(R)$ (black, solid) along with the corresponding eigenstates $\xi_i^{(1)}$ (colored and indicated by a symbol). For the state analysis, we consider the probability of finding the system in the eigenstates $\xi_i^{(1)}$ irrespective of the angle $\theta$. This is termed the fidelity $F_i(t)=\sum_j\left|\braket{\xi_i^{(1)}|\langle\varphi_j^{(2)}|\chi}\right|^2$. Figures \ref{fig:perio} (b)-(c) show the fidelities with matching symbols and colors according to (a). We find that the number of states participating in the overall dynamics is very sensitive to the waiting time between quenches $T$. In the case (b), where $T=3\,\mathrm{ns}$, many states, also highly excited ones, contribute. This corresponds to a delocalization of the wave packet across several potential wells, e.g.~$\xi_6^{(1)}$ shown in figure \ref{fig:perio} with a purple square participates in the dynamics and is localized in the outer well, which is not initially populated. For $T=5\,\mathrm{ns}$ (c) only two states dominate the evolution: The initial state $\xi_3^{(1)}$ and the energetically adjacent state $\xi_4^{(1)}$ that is localized to the adjacent well and exhibits two radial nodes. After propagation until $t=50\,\mathrm{ns}$, equal population of the two states $\xi_3^{(1)}$ and $\xi_4^{(1)}$ is achieved, which means that radially, the molecule exists in a superposition of states that are localized to distinct potential wells. At this point in time, we discontinue the quenches and observe the field-free evolution: The molecule coherently oscillates back and forth between the two states (see figure \ref{fig:perio} (c)).
			
\section{Conclusion} \label{sec:conc}

	We unravel the dynamical behavior of trilobite Rydberg molecules exposed to a homogeneous electric field. The analysis of the electronic structure reveals that the multi-well potential landscape is tilted by the electric field depending on the angle $\theta$ between the external field and the internuclear axis. For propagation protocols, in which the molecule is initially prepared in a field-free configuration and dynamically evolves in the presence of the field, the quenching of the field leads to intra-well oscillations of the radial density and oscillations around the equilibrium angle $\theta=\pi$ of the angular density. When the molecule is initially prepared as the ground state in the presence of the field, i.e.~localized in $\theta$, and the field is quenched to the opposite direction, the molecule starts to rotate with a field-dependent frequency. The radial intra-well oscillations are accompanied by large-amplitude inter-well oscillations in strong fields $F>325\,\mathrm{V/m}$, when the radial position of the ground state has shifted to the adjacent outward potential well. Furthermore, we explore the possibility to prepare the molecule in a superposition of different radial configurations localizing in distinct potential wells by a series of periodic quenches with varying frequency.
	
	We expect our results to be readily testable in state-of-art experiments, which have observed vibrational eigenstates of trilobite molecules after photoassociation from ultracold atomic gases and subsequent ionization. The internuclear separation is experimentally accessible via measurement of the permanent electric dipole moment. After photoassociation of the molecules, the dynamics can be triggered by a fast switch of the external electric field and the radial separation can be probed time dependently to reconstruct the wave packet propagation. The angular structure can be resolved, employing the recently developed quantum gas microscopes \cite{Veit2020,Geppert2020}.
	
	A natural extension of this work would be to allow for continuous changes of the electric field enabling state engineering by optimal control. Molecular dynamics in a magnetic field is equally of interest, a major drawback in practice being the larger switching times. However, combined electric and magnetic fields might allow for superpositions of different angular configurations. This study focuses on molecular dynamics on a single PES. Within our employed method, it is straightforward to extend the setup to a number of coupled PES and investigate dynamics beyond the Born-Oppenheimer approximation. A prime example would be the coupled PES of the trilobite and butterfly molecules. In our model, we neglect $p$-wave interactions, which introduces only minor modifications to the overall structure of the trilobite. However, the additionally arising butterfly PES provides a decay mechanism of the trilobite molecule for associative ionization, which can additionally be tuned by a weak external field. This establishes an interesting avenue to study chemical reactions of ultra-long-range Rydberg molecules in unprecedented detail.
		
\begin{acknowledgements}
	FH thanks Matthew Eiles for fruitful discussions, especially his input on the electronic structure. KK gratefully acknowledges a scholarship of the Studienstiftung des deutschen Volkes. PS acknowledges support from the Deutsche Forschungsgemeinschaft (DFG) within the priority program "Giant Interactions in Rydberg Systems" [DFG SPP 1929 GiRyd project SCHM 885/30-1].
\end{acknowledgements}

\bibliographystyle{apsrev4-1}
%

\end{document}